# Perception of Digital Privacy Protection: An Empirical Study using GDPR Framework

**Full research paper**

**Hamoud Alhazmi**

Department of Computer Science and Engineering

The Ohio State University

Columbus, OH 43210, USA

Email: alhazmi.8@osu.edu

**Ahmed Imran**

Faculty of Science & Technology

University of Canberra

Canberra, ACT 2617, Australia

Email: Ahmed.Imran@canberra.edu.au

**Mohammad Abu Alsheikh**

Faculty of Science & Technology

University of Canberra

Canberra, ACT 2617, Australia

Email: mohammad.abualsheikh@canberra.edu.au

## Abstract

Perception of privacy is a contested concept, which is also evolving along with the rapid proliferation and expansion of technological advancements. Information systems (IS) applications incorporate various sensing infrastructures, high-speed networks, and computing components that enable pervasive data collection about people. Any digital privacy breach within such systems can result in harmful and far-reaching impacts on individuals and societies. Accordingly, IS organisations have a legal and ethical responsibility to respect and protect individuals' digital privacy rights. This study investigates people's perception of digital privacy protection of government data using the General Data Protection Regulation (GDPR) framework. Findings suggest a dichotomy of perception in protecting people's privacy rights. For example, people perceive the right to be informed as the most respected and protected in Information Technology (IT) systems. On the contrary, the right to object by granting and withdrawing consent is perceived as the least protected. Second, the study shows evidence of a social dilemma in people's perception of digital privacy based on their context and culture.

# 1   Introduction

While privacy is considered a fundamental human right in most Western countries, the understanding of the privacy concept varies considerably among people based on individual life experiences and knowledge (Goel et al. 2019; Moshki and Barki 2014). In information systems (IS) research, the privacy concept includes concerns, interests, and functions Bélanger and Crossler (2011). While as a concept, privacy has been studied extensively in information systems, perception of privacy, and privacy rights, particularly by different socio- cultural demographics, attracted limited attention (Redmiles et al. 2017). Also, it is treated as a complex concept that can be viewed and studied from multiple perspectives (Pavlou 2011).

Privacy of data, particularly related to Internet use, became a significant concern in the last decade (Hong and Thong, 2013). Lately, privacy concerns surrounding data collection and storage have turned into a major barrier to the adoption of various Internet of Things (IoT) devices and wearables, which are different from those of Internet-based online transactions (Paul et al. 2020). A few studies have also identified privacy concerns as a barrier to the adoption of the IoT. For instance, Coughlan et al. (2012) studied the factors affecting the acceptance of home-based IoT technologies, and Prasad et al. (2012) attempted to understand what influences the information-sharing preferences and behaviour of users of mobile health devices.

While studies have recognised privacy issues as a hurdle, they have not delved into the specific worries of users or the factors that influence these concerns. Calls for such research using a broader diversity of sampling populations were solicited so that appropriate information systems for the protection of privacy could be developed (Baruh et al. 2017; Bélanger and Crossler 2011; Bélanger and James 2020). Against this backdrop, this paper attempts to fill the gap by offering an alternate perspective through an empirical study, which is likely to add some new insight into the IS discourse on privacy.

Modern IS solutions include components that enable data collection, sensing, transmission, storage, and processing. Data lies at the core of any information system, enabling real-time optimization, revealing critical trends, driving informed decisions, and drawing valuable insights. Examples of user data in IS solutions include biometrics, healthcare records, geolocation coordinates, shopping behaviour, selfies, fitness and daily routines, and text messages. Therefore, users have genuine concerns about the possibility of digital privacy breaches while using IS and data management services.

Digital privacy breaches have acute and far-reaching impacts on people and societies, where success of government IT service (eGovernment) delivery has often been tied to overcoming issues with privacy and security (Imran et al. 2013). Therefore, several technical solutions have been proposed to address the digital privacy problem, including distributed authentication protocols (Shin et al. 2021) and intrusion detectors in zero-touch networks (Naeem et al. 2023). However, digital privacy cannot be fully preserved without proper governing regulations (Hadzovic et al. 2023). Accordingly, governments and legislative bodies have introduced several digital privacy standards and guidelines to govern all data-related operations. The GDPR (European Parliament and Council of the European Union 2016) provides a set of digital privacy standards for governing how, when, why, and what data can be collected about users. The GDPR defines the digital privacy rights of users, shielding all aspects of data collection, transmission, storage, and processing in IS solutions. The term "digital privacy rights" in accordance to GDPR (2016) refers to the specific rights granted to individuals regarding the protection and control of their personal data in the digital environment. Previous research has shown that applying the GDPR guidelines moderately improves the attained digital privacy for end users (Momen et al. 2019).

This paper studies people's perception of digital privacy protection using the GDPR as a reference benchmark. We collected survey responses from 776 respondents to study how individuals perceive the levels of digital privacy protection that their governments provide when accessing IS solutions. We first operationalize the digital privacy protection concept based on the state-of-the-art privacy rights of the GDPR. Accordingly, we designed an online survey study with ten questions that enable the exploration of two research areas. Research area A includes eight questions (Q1-8) to measure the digital privacy levels provided by governments. Research area B consists of two questions (Q9 and Q10) that address the social dilemma in digital privacy, e.g., how people perceive the importance of their privacy and the privacy protection of others. We also designed three attention checks (A1-A3) and three timing checks (T1-T3) and applied rigorous statistical analysis to verify the research study's quality and ensure the reliability and validity of the collected responses.

The research study reveals three significant results. First, a disparity exists in protecting the digital privacy rights of users. For example, people perceive the right to be informed (Q1) as the most respected and protected in IS solutions. On the contrary, the right to object by granting and withdrawing consent (Q7) is perceived as the least protected. Second, rigorous statistical analysis reveals that the research study meets the consistent reliability measures and high confidence intervals. Third, the research study reveals a social dilemma in digital privacy. In particular, 82% of users think governments should enforce stringent privacy regulations on domestic companies and organizations that collect their personal data in all cases, regardless of the data collection's economic, employment, and crime prevention benefits (Q10). On the contrary, 70% of users think governments should allow domestic companies and organizations to collect personal data on individuals residing outside their countries if economic, employment, and crime prevention benefits are anticipated (Q9).

The rest of this paper is organized as follows. We first discuss related works followed by an overview of digital privacy and privacy rights depicted in the GDPR. Then, we discuss how digital privacy must be met in all IS-enabling technologies, including the IoT for data sensing, networking for data transmission, cloud computing for data processing, and machine learning and data analytics. After that, we discuss the research methodology and the operationalization of the digital privacy protection concept into abstract privacy concepts, which are measured using the research study. Then, we present the results and statistical analysis of the research study, which shows the disparity and social dilemma in digital privacy. We also suggest recommendations for mitigating the disparity and social dilemma problems in digital privacy protection. Finally, we outline important future research directions and conclude the paper.

## 2   Related Work

In this section, we will focus on prior studies that researched privacy concerns and provide a foundational understanding of how individuals perceive threats to their personal information.

The concept of privacy holds varied interpretations across disciplines: as a right or entitlement in legal literature, as a state of restricted access or isolation in social psychology, and as control over information in information systems research (Pavlou 2011). Pavlou (2011, p.977) defines privacy as the individual's right to determine what personal information should be shared with others and in what circumstances. This involves managing how personal data is obtained and utilised, which enables individuals to create and maintain their self-identities within different social settings, giving them the freedom to control what, how, and to whom they share information about themselves (Proshansky et al. 1970; Stone et al. 1983; Warren and Brandeis 1989).

Data or information privacy is a subset of privacy where people attach substantial value and "acts like a shield for one's personal identity" (Floridi 2006, p. 111). Individual privacy concerns are focused on collection, control, and awareness of privacy practices (Malhotra et al. 2004). Privacy concerns represent overarching worries stemming from individuals' inherent fear of potential information privacy breaches (Malhotra et al. 2004). However, consumers willingly share their private information because they perceive privacy as a commodity that can be traded for economic gains (Bennett 1995). Thus the balance between the risks and benefits of information privacy is a contested issue with significant theoretical and practical implications (Pavlou 2011).

Antecedent-privacy concern-outcome (APCO) developed by (Smith et al. 2011) has been recognized as a prominent framework that focuses on factors that lead to privacy concerns. According to Paul et al. (2020), privacy concerns are influenced by intrinsic factors—disposition to value privacy and extrinsic factors— privacy policies and the GDPR. Their study revealed that the GDPR lowers users' average privacy concerns. Furthermore, a heightened perception of the effectiveness of privacy policies decreases perceptions of privacy risks and enhances perceptions of privacy control.

Dinev and Hart (2006) posit privacy concerns involve an individual's anxiety regarding the potential loss of privacy due to the willing or unwilling revelation of personal information. Goel et al. (2020) attempted to understand the diverse ways individuals value privacy, contending that although people often treat privacy as a singular concept, their privacy-related actions are influenced by specific situational factors. Individuals perceive privacy differently depending on the circumstances and prioritise their decisions accordingly. Tewari and Mills (2021) saw the relationship between privacy concerns, information privacy awareness (legal frameworks) and identity threats in the context of digital identities.

While user concerns regarding privacy are extensively discussed in the literature, there is limited research exploring the origins of these concerns (Bakke 2005; Jozani et al. 2020), and there remains a gap in understanding the factors influencing Privacy Concerns Related to IT (PCIT), their development, and the connection between general PCIT and PCIT concerning particular technologies (Moshki and Barki,2014). As Kim and Bhagat (2022) insist, there is still a lack of research on other possible constructs surrounding privacy concerns.

Tronnier and Biker (2022) examined citizens' privacy concerns regarding the digital euro and explored the dimensions of privacy concerns citizen's exhibit. Their findings, however, demonstrate the importance of factors such as culture and demographics, with other antecedents and dimensions being less important. Han and Ellis (2016) adopt the IS continuance model to study online users' continuance intentions toward privacy-protection practices and posit that online users' inconsistent behaviours may result from the inappropriate designs of the practices and users' unwillingness to continue to use the practices. Xu et al. (2011), in their study, drawing on the Communication Privacy Management (CPM) lens, developed a research model suggesting that an individual's privacy concerns form through a cognitive process involving perceived privacy risk, privacy control, and their disposition to value privacy.

Bakke (2005) proposes the Privacy Invasiveness Perception Scale (PIPS) to measure the impact of information technology characteristics on users' privacy perceptions. This scale is expected to enable organizations to choose more adoptable information technologies that minimize user resistance based on individual privacy concerns. Pentina et al. (2016) investigated the impact of personality and cross-cultural variances on the privacy calculation model. It revealed that individuals with high sociability traits are more inclined to share information, while those scoring high in neuroticism tend to prioritize keeping information private.

Goel et al. (2019) found that individuals may choose to divulge private information not solely due to situational factors but also because their concerns about privacy in a specific domain are not as pronounced as their overall privacy concerns. It is thus imperative to identify the dispositional differences among the distinct groups of people and to better understand the reasons for this dichotomy of privacy beliefs between these groups (Goel et al. 2020). Hanneke et al. (2023) studied how different aspects of privacy management systems affect customer choices in GDPR-regulated systems. It was found that users tend to be more inclined to share their data in exchange for marketing incentives and discounts.

## 3 Digital Privacy Rights—Overview

This section argues the significance of data in IS solutions, followed by a discussion of the eight GDPR privacy rights.

### 3.1 Data as the fuel of IS solutions

Digital privacy signifies the ability of users to control the collection, transmission, storage, and processing of their data (European Parliament and Council of the European Union 2016). IS solutions and services use the IoT for data sensing, high-speed networks for data transmission, cloud computing for data storage and computing, and machine learning for data analytics and automated decisions. Figure 1 shows the architecture of modern IS solutions that collect, transmit, store, and process users' data. IoT devices are embedded with various sensors, e.g., GPS, cameras, accelerometers, and microphones, which sense data about people and their surroundings. The data is transmitted through high-speed wireless, e.g., 5G, and wired, e.g., fibre optics, networks to service providers. Service providers use efficient cloud computing, e.g., Amazon Web Services, to store and perform computations on the data. Machine learning is applied to learn the relation- ships in users' data and create new services, e.g., real-time traffic monitoring, speech recognition, and fitness trackers.

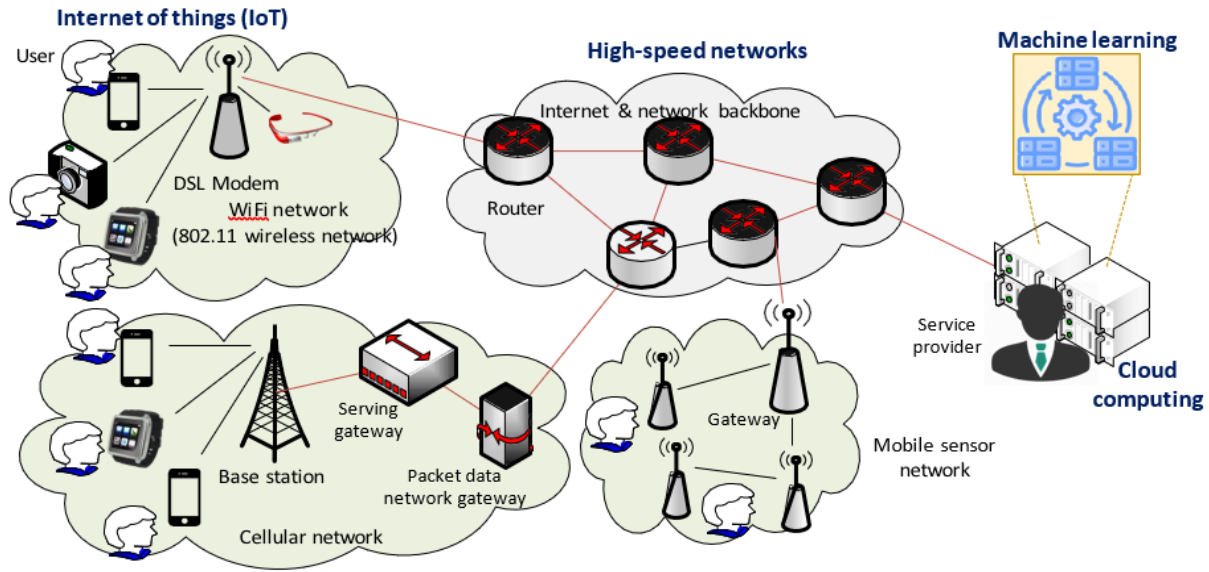

*Figure 1: IS solutions collect, transmit, store, and process massive user data*

## 3.2   GDPR privacy rights

The GDPR (European Parliament and Council of the European Union 2016) includes well-defined and state- of-the-art regulations to preserve digital privacy rights for individuals. In particular, the GDPR privacy rights explain what service providers must provide to users before performing data collection, transmission, storage, or processing.

- Right to be informed: The right of users to know who is collecting and accessing their data.
- Right of access: The right of users to access their data.
- Right to rectification: The right of users to request correcting inaccurate records of their data.
- Right to erasure: The right of users to be forgotten by deleting their data and preventing future data collection.
- Right to restrict processing: The right of users to restrict the processing of specific categories of their data.
- Right to data portability: The right of users to transfer their data to specific recipients of their choice.
- Right to object: Users retain the right to provide and withdraw consent on their data processing and collection.
- Rights concerning automated decision-making and profiling: Users' right to opt-out from using their data in automated systems, including machine learning and artificial intelligence (AI).

Complying with the GDPR requirements can significantly impact IS solutions and services, including market competition (Wu and Pang, 2021), data sharing (Hanneke et al. 2023), and financial operations (Kircher and Foerderer, 2021). Consequently, investigating digital privacy perceptions through the lens of the GDPR presents an important research problem.

## 4   Research Study—Design and Methodology

Developing a research study to identify and measure compliance with digital privacy regulations is not simple. Self-reported responses to research studies with questionnaires have limitations, including biased over-reporting, under-reporting, misunderstanding questions, and social desirability, i.e., providing favorable responses by the social group. These challenges are tackled by following well-established, rigorous survey design processes, e.g., operationalization, validation, pilot studies, and quality checks (Red- miles et al. 2017; Saris and Gallhofer 2014). This section discusses the process of operationalizing digital privacy in IS solutions into accurate survey questions. This section first defines the operationalization links between digital privacy protection and the fundamental GDPR privacy rights. Then, it discusses the bare assertions used to represent digital privacy levels in our questionnaire.

## 4.1 Research questions

This study applies the GDPR framework to examine critical issues surrounding digital privacy in the context of IS solutions. The study is guided by the following research questions:

- How do people perceive the protection of their digital privacy rights by their government when it collects and processes personal data?
- Which digital privacy rights are perceived as most and least respected and protected by government data collection practices?
- Is there evidence of a social dilemma in people's perceptions of digital privacy?

The paper aims to provide a comprehensive understanding of public perceptions of digital privacy in IS solutions, identifying key areas of concern and potential areas for policy improvement. It also aims to identify which of digital privacy rights the public perceives as being well-protected and which are seen as vulnerable or inadequately addressed. Moreover, this study investigates whether there is a conflict between individual and collective interests in digital privacy protection. A social dilemma might involve situations where individuals prioritize personal benefits, e.g., convenience and finance, over privacy protections of others.

## 4.2 Choice of operationalization

Digital privacy protection is a complex construct that is difficult to measure. We therefore applied the operationalization method of defining the concepts-by-postulation (complex variables) of a study using concepts-by-intuition (abstract, measurable variables) (Saris and Gallhofer, 2014). Figure 2 shows the operationalization of the digital privacy protection concept into the corresponding privacy concepts and the developed questionnaire questions that represent the key concepts of the study.

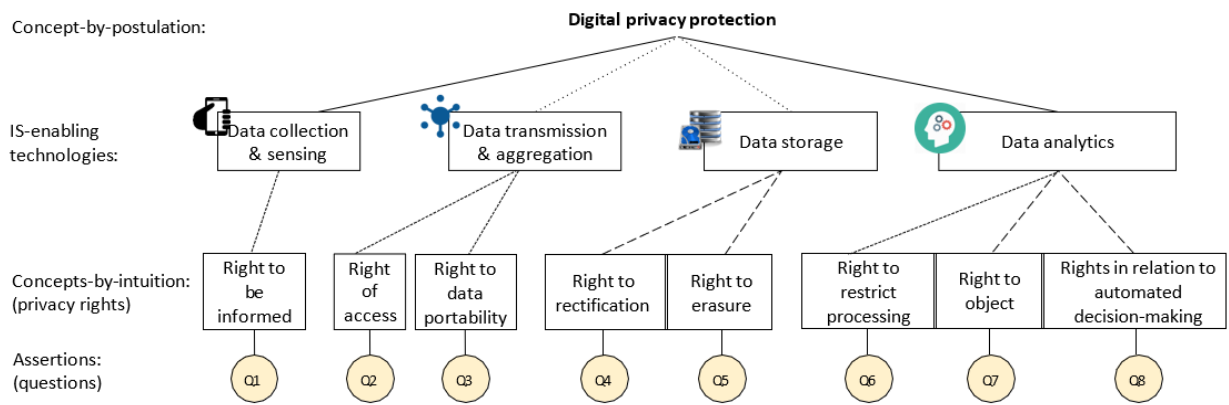

*Figure 2: Digital privacy in IS is systematically measured with Q1-8 in the online survey study*

The privacy rights are abstract concepts, and they can be clearly measured with questions. The questionnaire captures all functional components of a modern IS solution, including data collection and sensing (IoT and crowdsensing), data transmission and aggregation (networking), data storage (cloud computing), and data analytics (machine learning). The corresponding questions that measure all privacy rights are as follows.

- **Q1:** I receive clear information on how my government collects my personal data, including who is accessing and processing the data and the data collection purposes.
- **Q2:** I can access copies of my personal data, which my government has collected.
- **Q3:** I can transfer my personal data, which my government has collected, to third-party recipients, e.g., organizations, of my choice.
- **Q4:** I can correct my personal data, which my government has collected, when it contains inaccurate, invalid, or misleading data.

- **Q5:** I can request deleting specific records of my personal data, which has been collected by my government when the data is no longer needed for the original purpose.
- **Q6:** I have the option and control to restrict the processing of specific categories of my personal data, which my government has already collected.
- **Q7:** I have the control and ability to grant or withdraw consent on collecting and processing my personal data by my government at any time.
- **Q8:** I have the option and control to opt-out from using my personal data, which my government has collected in making decisions and profiling, based solely on automated processing.

Two questions (Q9 and Q10) are also included to measure the social dilemma in privacy preservation.

- **Q9:** My government should allow domestic companies and organizations to collect personal data on individuals residing outside my country if there are economic, employment, and crime prevention benefits for my country.
- **Q10:** My government should enforce strict privacy policies on domestic companies and organizations that collect my data in all cases, regardless of the economic, employment, and crime prevention benefits of the data collection.

### 4.3  Reliability and validity of the instruments

Considering the reliability and validity of survey data is critical to delivering credible results, we applied three-step quality checks and filtering on the collected responses. Two of these are attention checks, and one is a timing check.

- Attention checks: A1 and A2 are attention checks with feedback, i.e., they warn the respondent of in- correct responses and describe the importance of accurate responses. A3 is an attention check without feedback, i.e., it stores the answers provided by the respondents without providing explicit feedback on wrong answers. "Is it correct to say that 2 plus 3 equals 1?" is an example of the used attention checks. 14.28% of the respondents failed A3. This fail percentage is slightly higher than some previous studies, e.g., 10% of the respondents failed similar simple attention checks in (Bonnefon et al. 2016).
- Time checks: Several studies have discussed the relationship between speeding and response quality, e.g., straight-lining, in web questionnaires (Zhang and Conrad, 2014). We implemented three timers to compute the time needed to answer Q1-8, Q9, and Q10, respectively. We excluded all responses that fall within the first percentile of timing.

A pilot test was conducted among more than ten local respondents to eliminate the ambiguity of survey questions. Feedback from those respondents was taken to improve the wordings, remove ambiguity, and change the order and structure of the questions.

## 5  Results and Analysis

This section discusses the results and statistical analysis of the research study. First, the disparity problem in digital privacy protection is presented by analyzing the users' responses. Second, the social dilemma of digital privacy protection is discussed.

### 5.1  Data collection

We recruited 776 respondents using Amazon Mechanical Turk (MTurk) and manual sharing. We restricted the participation to respondents in the United States, Germany, Bangladesh, and India. We could not collect sufficient responses from Bangladesh using MTurk, so we collected more than 100 responses by circulating the online survey study with our network in Bangladesh, i.e., manual sharing. This is a diverse data collection compared to most previous survey studies, which collect responses from one country (Bonnefon et al. 2016; Redmiles et al. 2017). Previous works have shown that responses collected using MTurk in security and privacy studies provide a good representation of the general population (Redmiles et al. 2019).

## 5.2 Digital privacy disparity

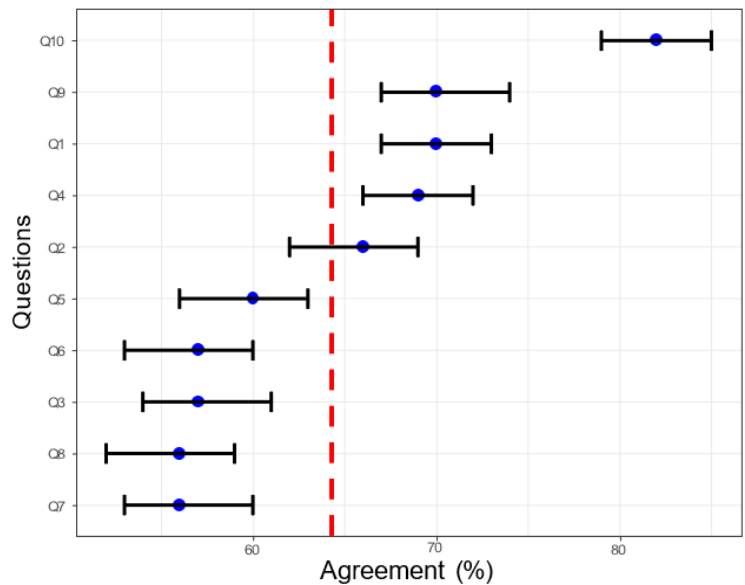

*Figure 3: Agreement percentages to questions Q1-10 and their 95% confidence intervals*

Figure 3 shows the percentage of agreement responses, i.e., strongly agree and agree, to Q1-10 and their corresponding 95% confidence intervals (CI). 95% CIs are the range of values that contain the actual agreement percentage of the questions—only a 5% chance that the confidence intervals may exclude the actual agreement percentage of the population if the research study is repeated (1 time out of 20 repetitions).

It can be noted from Figure 3 that a disparity exists in protecting the privacy rights of individuals. The agreement percentages, in descending order, are 56% for Q7 (right to object) with CI of [53,60], 56% for Q8 (rights in relation to automated decision-making) with CI of [52,59], 57% for Q3 (right to data portability) with CI of [54,61], 57% for Q6 (right to restrict processing) with CI of [53,60], 60% for Q5 (right to erasure) with CI of [56,63], 66% for Q2 (right of access) with CI of [62,69], 69% for Q4 (right to rectification) with CI of [66,72], and 70% for Q1 (right to be informed) with CI of [67,73]. These results reflect people's perception that the right to be informed is more respected and protected than the remaining privacy rights. On the contrary, the right to object and withdraw consent is the least protected. The red-coloured vertical line represents the average agreement percentage for all questions.

## 5.3 Statistical analysis

A correlation matrix can be created by computing the correlation coefficients between Q1-10 to gain insights into the relationships among the questionnaire questions. A high correlation value (close to one) suggests that questions measure a similar underlying construct. In contrast, a low correlation value (close to zero) implies that they measure different constructs. Figure 4 shows the correlation among the research questions Q1-10. It can be noted that the research questions Q1-8 are highly correlated, which is expected as Q1-8 address different aspects of digital privacy, e.g., different aspects of one concept. Q9 and Q10 are loosely correlated with other questions as they measure the social dilemma problem.

Cronbach's α is a metric that indicates the internal consistency (or reliability) of a set of questions that measure a single latent construct or underlying variable. An adequate internal consistency can be concluded when α is greater than 0.70 (Morera and Stokes, 2016). In this study, Q1-8 (a set of question items) measure digital privacy protection (the single latent construct). Our statistical analysis indicates that this research study is reliable with internal consistency reliability of α = 0.92 for Q1-8. This is expected as Q1-8 measures closely related rights in digital privacy.

These statistical results indicate a high level of reliability for the questions measuring the digital privacy protections provided to users and reinforce the high quality of the data collected and study design.

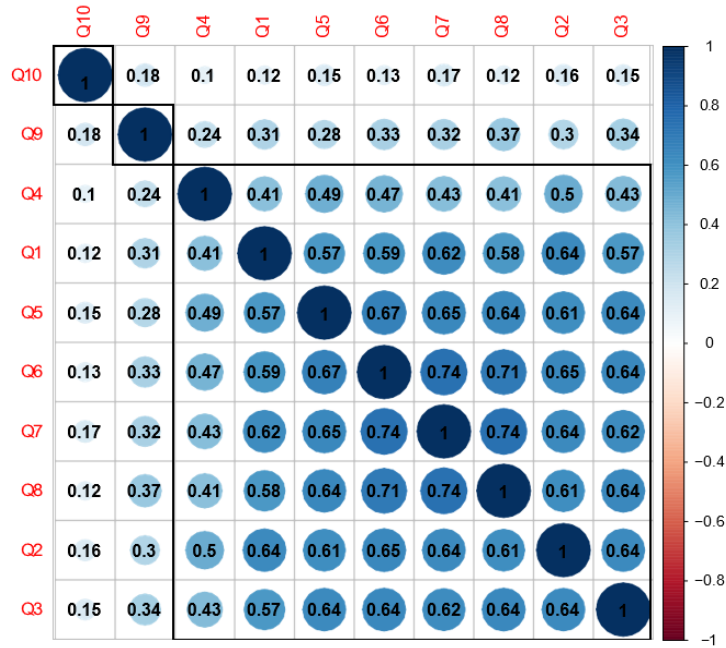

*Figure 4: The correlation matrix of Q1-10 shows that Q1-8 correlate closely*

### 5.4 Social dilemma of digital privacy

A social dilemma arises when people have conflicting incentives between driving their self-interest and the collective interest of their community. For example, the "tragedy of the commons" is a known social dilemma that happens when individuals act in their self-interest and deplete shared common-pool resources, even when it is detrimental to the collective group (Hardin 1968).

As shown in Figure 3, 82% of the respondents think governments should enforce strict privacy policies on domestic companies and organizations that collect their data in all cases, regardless of the data collection's economic, employment, and crime prevention benefits (Q10). On the contrary, 70% of respondents think governments should allow domestic companies and organizations to collect personal data on individuals residing outside their countries if there are economic, employment, and crime prevention benefits (Q9).

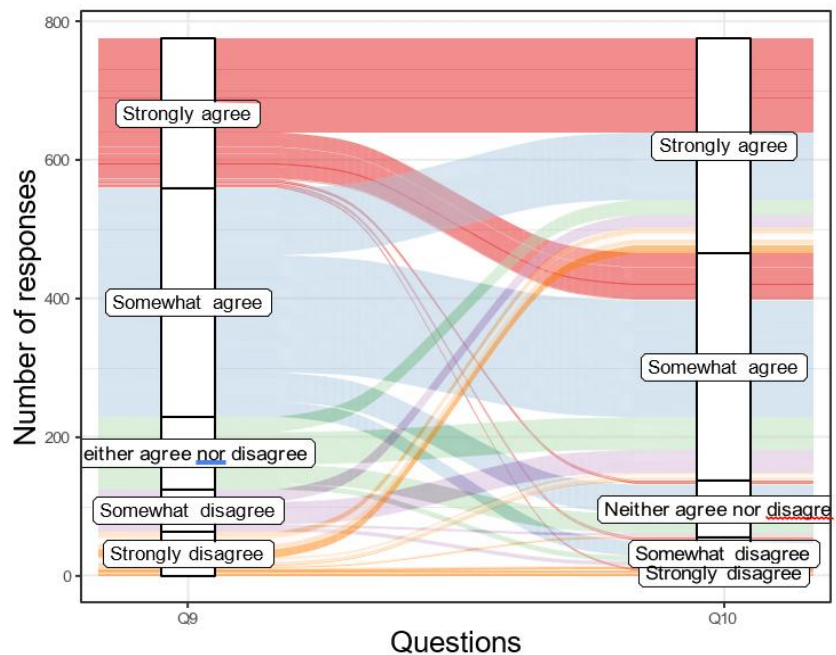

*Figure 5: Social dilemma between personal privacy (Q10) and the privacy of others (Q9).*

Figure 5 is an alluvial diagram that shows the changes in responses to Q9 (other privacy) and Q10 (individual privacy). It can be noted that more people provided "strongly agree" responses for Q10, which reflects the high importance of individual privacy. It can be noted that most respondents support data collection about individuals residing outside their countries if there are economic, employment, and crime prevention benefits.

# 6 Recommendations for IS Stakeholders

The disparity problem and social dilemma in privacy protection reinforce social injustice. This section provides recommendations for mitigating the disparity and social dilemma problems in digital privacy with three main dimensions (educational, legislative, and technical extents).

## 6.1 Education and awareness initiatives on privacy rights

Many IS users may not be fully aware of the pervasive nature of existing data sensing and processing technologies. Previous research has shown that users may not use privacy tools when they lack a clear understanding of their benefits and architectures (mental models) (Renaud et al. 2014). Therefore, educational interventions, awareness programs, and campaigns must be initiated to increase citizens' awareness of privacy protection and privacy rights. In addition, online services provided by governments must include an explicit description of the privacy rights.

Knowledge and awareness on disclosures, marketing materials, loopholes related to online privacy on social networking sites, malware, and adware to collect personal information about the users should be widely disseminated to prevent possible breaches (Grama 2020).

## 6.2 Unified privacy regulations worldwide

There are various legislation policies all over the world, e.g., the GDPR (European Union), the Privacy Act 1988 (Australia), the Personal Information Protection Law (China), the California Consumer Privacy Act (California, USA), and the General Personal Data Protection Law (Brazil). The differences in privacy legislation laws result in various interpretations of privacy rights and the acceptable levels of privacy protection. A multi-stakeholder view of privacy, capturing local contextual and cultural interpretations of privacy, should be reflected in the policy. Questions should be raised about to what extent the government should invade its citizens' privacy, showing the pretext of national security.

Potential boundaries and trade-offs between individual rights and security must be drawn carefully by examining their short- and long-term implications for individuals and society. Associated laws should keep up with the fast-paced technological development changes to prevent possible gaps and confusion. To ensure that privacy protections are robust enough to manage future innovations in data processing and analytics, legislation may aim for technical neutrality, thereby avoiding the need for frequent legal updates (Rommetveit and Van Dijk, 2022).

## 6.3 Accessible privacy tools

A critical aspect of mitigating the disparity and social dilemma problems in digital privacy protection is providing accessible and easy-to-use privacy tools for both users and service providers. Service providers need IS tools to meet the privacy rights in IoT, networking, cloud computing, and machine learning. Simultaneously, users require usable tools for verifying the service providers' compliance with the privacy regulation (Tanuwidjaja et al. 2020).

# 7 Future Research Directions

This section suggests fundamental research directions in IS digital privacy that can be pursued in the future.

## 7.1 Socio-demographic and socio-economic status

People from different socio-economic backgrounds can experience different security and privacy problems and have different expectations about their privacy-related issues (Redmiles et al. 2017). For example, some people are more willing to share their data to gain access to promotion and discount materials. Such perception of privacy is also likely to result from locally situated work practices and socially negotiated realities that impact privacy behaviour in those settings. Future research can use the theoretical approach of "situating culture" (Weisinger and Trauth, 2002) to better understand contextualism and cross-cultural issues in privacy perception among people of different countries and regions.

### 7.2 Societal issues and individual digital privacy

An important research direction is perhaps studying the impact of major societal issues on individual digital privacy. For example, contact tracing applications have been implemented by many governments during the COVID- 19 disease outbreak to monitor interactions with infected patients. Even though tracing applications are essential for monitoring infection trends, tracing methods enclose substantial privacy risks (Tedeschi et al. 2021). An interesting future research direction could be analyzing the impact of societal issues, e.g., disease outbreaks, on the acceptance of higher privacy risks and the trade-offs between privacy compromises and potential benefits. In addition, digital privacy regulations need to be amended to specify the conditions for data tracing when addressing major societal issues.

### 7.3 Mixed method study of digital privacy inequality

A mixed-method study followed by a qualitative inquiry should be conducted to provide a deeper understanding and rich insight into complex socio-cultural and socio-political dimensions influencing the perception of privacy and to complement quantitative survey data. Such an approach helps explain the underlying causes of change in citizens' behaviour over time, the role and impact of power relations between governments and citizens, and trust in IS solutions and services.

## 8 Conclusion

This paper has offered additional insight and an alternative approach to people's perception of digital privacy protection of government information using the GDPR as a reference benchmark. We have found a disparity in protecting privacy rights of individuals in IS solutions. Our findings show the evidence of a social dilemma in digital privacy. Most respondents are concerned about their digital privacy while supporting data collection about individuals residing outside their countries if economic, employment, and crime prevention benefits are anticipated. The study offers interesting dimensions for future research directions and provide recommendations to address the disparity and social dilemma problems.

Digital privacy breaches pose significant threats to individual freedom and dignity in the modern world, which has substantial ramifications for creating a just society. Due to the increasingly global nature of digital access and use, a global consensus must be reached to preserve human dignity and privacy rights for safe societies.

# Copyright